# The Great Eruption of η Carinae

Kris Davidson and Roberta Humphreys

Minnesota Institute for Astrophysics, University of Minnesota

During the years 1838-1858, the very massive star η Carinae became the prototype supernova impostor: it released nearly as much light as a supernova explosion and shed an impressive amount of mass, but survived as a star.[1] Based on a light-echo spectrum of that event, Rest et al.[2] conclude that "a new physical mechanism" is required to explain it, because the gas outflow appears cooler than theoretical expectations. Here we note that (1) theory predicted a substantially lower temperature than they quoted, and (2) their inferred observational value is quite uncertain. Therefore, analyses so far do not reveal any significant contradiction between the observed spectrum and most previous discussions of the Great Eruption and its physics.

Rest et al. state that a temperature of 7000 K was expected, and that 5000 K is observed. These refer to outflow zones that produced most of the emergent radiation. For the 7000 K value those authors cite a 1987 analysis by one of us,[3] but they quote only a remark in the text, not the actual calculated values. According to Figure 1 of the 1987 paper, η Car's Great Eruption should have had a characteristic radiation temperature in the range 5400—6500 K, not 7000 K. (Here we assume mass loss > 1 $M_{sun}$ per year and luminosity > $10^7$ $L_{sun}$.[1]) Text references to 7000 K in 1987 concerned less extravagant outbursts, and η Car was explicitly stated to differ from them.  Moreover, to establish a conflict between observations and expectations, new calculations with modernized opacities would be needed.

The circa-5000 K "observed" temperature described by Rest et al. is based on a derived classification for the light-echo spectrum, using automated cross-correlations with a set of normal supergiant stars. This technique may be suitable for mass-production normal spectra, but any non-routine object requires specific feature-by-feature comparisons instead.  One of the first principles of stellar classification is to separate luminosity from temperature criteria, but all the reference stars in this case were far less luminous than η Car's eruption.  (Luminosity correlates with surface gravity, which affects gas density and thereby the spectrum.)  Furthermore, emission lines appear to be present and may contaminate an automated analysis;  but without access to the spectrum we cannot verify this. Rest et al. used a temperature calibration from a 1984 reference[4] taken from an even older publication in 1977.[5]  Considerable work has been done since then, and for the highest luminosities, each spectral type has a substantial range of possible temperatures – e.g. 5100—6200 K for the G2—G5 spectral types favored in their paper.[6,7,8]  In view of all the above points, the temperature range indicated by stellar classification overlaps the theoretical expectations.

Moreover, η Car's eruption was a large-scale mass outflow, not a static atmosphere with definable surface gravity.  This distinction quantitatively alters the relation between absorption lines and the underlying continuum.  "Characteristic radiation temperature" in the 1987 theoretical description[3] is therefore defined differently from a normal star's "effective temperature." If spectral types are assigned to outflows, there is no reason to expect their temperatures to coincide with the stellar-atmosphere calibration adopted by Rest et al. This is not a question of stellar wind versus explosion; dense winds, stellar eruptions, and opaque explosions are basically alike in their emergent radiation physics,[1,3] and their density dependences $\rho(r)$ differ in character from normal stellar atmospheres.

In conclusion, so far as existing models allow anyone to say, *the observed spectrum appears consistent with what one expects for a giant eruption with η Car's parameters.*